# Experimental ratchet effect in superconducting films with periodic arrays of asymmetric potentials.

J. E. Villegas, E. M. Gonzalez and M. P. Gonzalez,
Departamento Física de Materiales, Facultad CC. Físicas, Universidad Complutense,
28040 Madrid, Spain.

J. V. Anguita
Instituto de Microelectrónica de Madrid, CSIC-CNM, Isaac Newton 8, P. T. M. Tres Cantos, 28760 Madrid, Spain

J. L. Vicent
Departamento Física de Materiales, Facultad CC. Físicas, Universidad Complutense,
28040 Madrid, Spain

Abstract.

A vortex lattice ratchet effect has been investigated in Nb films grown on arrays of nanometric Ni triangles, which induce periodic asymmetric pinning potentials. The vortex lattice motion yields a net dc-voltage when an ac driving current is applied to the sample and the vortex lattice moves through the field of asymmetric potentials. This ratchet effect is studied taking into account the array geometry, the temperature, the number of vortices per unit cell of the array and the applied ac currents.

Feynman used, in his Lectures on Physics,[1] a ratchet to show how anisotropy never could lead to net motion in an equilibrium system. Since then, asymmetric sawtooth potentials are called ratchet potentials and in general, a device with broken inversion symmetry is called a ratchet device. The ratchet effect occurs when asymmetric potentials induce outward particle flow under external fluctuations in the lack of any driving direct outward forces. The ratchet effect changes an ac source in a dc one. Ratchet effect spans from Nature phenomena to laboratory fabricated devices. In a ratchet, the energy necessary for net motion is provided by raising and lowering the barriers and wells, either via an external time-dependent modulation, for example an ac current injected in a superconducting film with asymmetric pinning centers,[2] or by energy input from a no equilibrium source, such as a chemical reaction, as for instance in biological motors.[3] During the last years, ratchet effect has called the attention of many researchers. A state of the art on the related topics Brownian motion and ratchet potential could be found in Reference 4.

The use of ratchet-like pinning potentials in superconductors has been the subject of theoretical approaches which deal with very different topics; for instance to remove flux trapped in superconducting devices,[5] fluxon optic,[6] logic devices,[7] etc. From the experimental point of view, some progress has been reported related to superconducting circuits,[8-10] and very recently in superconducting films with artificially fabricated arrays of asymmetric pinning centers.[2] In the present paper, we will address some of the properties of this superconducting ratchet effect. We will explore the dependence of the ratchet with the applied alternating current, the array shape, the temperature, and the number of vortices per array unit cell. We will show that periodic asymmetric potentials are crucial to produce the ratchet behavior, that the effect is enhanced decreasing the temperature, and finally, that the effect decreases when the applied magnetic field (number of vortices per unit cell of the array) increases.

The paper is organized as follows: First, we will summarize some results on the behavior of vortex lattice on artificially induced pinning potentials. After we will present the fabrication method and main characteristics of the films. Finally, the experimental ratchet effect results will be showed and discussed.

The dynamics of vortex lattice on artificially induced symmetric pinning potentials has received a lot of attention during last years. Different shapes of pinning centers and arrays, and different materials have been explored from the experimental as well as the theoretical points of view. As it has been shown,[11] the dc magnetoresistance of superconducting thin films with periodic arrays of pinning centers show minima when the vortex lattice matches the unit cell of the array. These minima are sharp (strong reduction of the dissipation) and equal spaced (two neighbor minima are always separated by the same magnetic field value). In the case of square arrays of nanostructured pinning centers, minima appear at applied magnetic fields

$H_m = n (\Phi_0/a^2)$, where $a$ is the lattice parameter of the square array and

$\Phi_0 = 2.07 \times 10^{-15}$ Wb is the magnetic flux quantum. Hence, the number of vortices $n$ per array unit cell can be known by simple inspection of the dc magnetoresistance R(H) curves, in which the first minimum corresponds to one vortex per unit cell, the second minimum to two vortices per unit cell, and so on. Moreover, the ratio between the dimension of the pinning center and the superconducting coherence length governs the maximum number of vortices that could be pinned in each one of the magnetic centers.[12] Therefore, we know, for selected values of the applied magnetic field, how many vortices are per unit cell and where they are, this is, if they are interstitial vortices or vortices in the pinning centers. Other interesting effects induced by these periodic potentials on the vortex lattice behavior are changes in the vortex lattice symmetry,[13] channeling effect in the vortex motion,[14] and effects related to the interplay between these artificial periodic pinning centers and the intrinsic random pinning centers of the sample.[15]

Electron beam lithography, magnetron sputtering and ion etching techniques are used to fabricate Nb thin films (200 nm thickness) grown on arrays of Ni triangles of thickness 40 nm. Si(100) is used as substrate. A cross-shape patterned bridge of 40 μm wide allows us injecting current into the sample. More fabrication and characterization details have been reported elsewhere.[2]

Fig. 1 shows the *dc* voltage drop $V_{dc}$ measured along the *x*-axis due to the *ac* current density $J = J_{ac}\sin(\omega t)$ injected along the *x*-axis in two samples A, Fig. 1(a), and B, Fig. 1(b), with different arrays and with one vortex per unit cell. Although the samples are quite different the experimental results are similar. In summary, the *ac* current density yields an *ac* Lorentz force on the vortices that is given by $\vec{F}_L = \vec{J} \times \vec{n}\phi_0$ (where $\vec{n}$ is a unitary vector parallel to the applied magnetic field), but the time averaged driving force on vortices is $\langle F_L \rangle = 0$. From the expression for the electric field $\vec{E} = \vec{B} \times \vec{v}$ (with *B* the applied magnetic field and $\vec{v}$ the vortex-lattice velocity), we get that the dc voltage drop $V_{dc}$ measured along the direction of the injected current is proportional to the average vortex-lattice velocity $\langle v \rangle$ in the direction of the ac driving force; in particular $V_{dc} = \langle v \rangle dB$ (where *d* is the distance between contacts and *B* the applied magnetic induction). For and *ac* current input (or *ac* driving force), the output is a non-zero *dc* voltage $V_{dc}$. This means that a net vortex lattice flow ($\langle v \rangle \neq 0$) arises from the *ac* driving force ($\langle F_L \rangle = 0$), what shows that the array of Ni triangles induces a ratchet potential landscape for the vortex lattice. The magnitude of $V_{dc}$ output depends on the amplitude of the *ac* current input $J_{ac}$: $V_{dc}$ peaks as a function of the *ac* input amplitude. This behavior agrees with results obtained from numerical simulations of vortex dynamics in asymmetric potentials.[16,17] As the *ac* amplitude increases, the net velocity increases monotonically from zero up to the maximum value. From this point the effect progressively smears out as stopping forces become negligible in comparison with the ac drive amplitude.

The inset of Fig 1(b) shows the same experiment but now the *dc* voltage drop $V_{dc}$ measured along the *y*-axis due to the *ac* current density $J = J_{ac}\sin(\omega t)$ injected along the *y*-axis. In sample A the *dc* signal is zero along the *y*-axis,[2] while sample B the dc voltage shows similar values than in the perpendicular direction [Fig. 1(b)]. The ratchet effect along *y*-axis vanishes in the A sample because the vortex lattice is then flowing on an array of symmetric periodic potentials. However, in sample B the pinning potential is also asymmetric along the *x*-axis, thus yielding a ratchet effect along this direction. To underlay that the asymmetric potential is the main ingredient of the ratchet effect, we have carried on a test injecting dc current. Fig. 2 shows the comparison between the *ac* ratchet effect, at the highest kHz frequency that is attainable in our experimental set up, and the pure *dc experiment*. Here the *ratchet signal* is extracted from *dc I(V)* curves: after a dc current is applied, first in the + *x*-axis $V(I_+)$, and then in the – *x*-axis $V(I_-)$, both curves are subtracted and the ratchet voltage is $V=V(I_+)-|V(I_-)|$. We can see from Fig. 2 that the *dc experiment* mimics the *ac* data, that is, the amplitude, shape and current window where the ratchet effect occurs with *ac* drive. Moreover, from Fig. 2 (b) one can clearly see that the ratchet effect is observed for *ac* (or *dc*) current amplitudes above the *dc* critical current. We can conclude that asymmetric potentials are the key ingredient of the ratchet behavior.

Fig. 3 shows the temperature dependence of the ratchet effect, in sample A, with the *ac* current injected parallel to the triangle basis (*x*-axis) and one vortex per pinning site. The temperature has several effects on the ratchet behavior: on one hand, the effect is enhanced, since the maximum rectification value $V_{dc}^{max}$ monotonically increases as temperature is decreased, and the shape of the peak gets steep. These trends are illustrated in the inset of Fig. 3. In this inset, the maximum in the rectification $V_{dc}^{max}$ is depicted as a function of the reduced temperature. As can be seen, $V_{dc}^{max}$ grows fast with decreasing temperature, and afterwards increases more slowly and finally seems to saturate. This behavior could be related to the *thermal noise*,[16] which assists vortex motion through fluctuations, and that washes out potential *asymmetry* at temperatures very close to $T_c$, strongly reducing the magnitude of the ratchet effect. Thermal noise disappears as temperature is lowered from $T_c$, what causes the fast increase in the rectification, while further temperature reduction has a less pronounced effect. On the other hand, the *ac* current amplitude at which the maximum in *dc* rectification voltage develops, $J_{ac}^{max}$, is shifted towards lower values as temperature decreases. This is also shown in the inset of Fig. 3. The origin of the shifting of the $V_{dc}$-$J_{ac}$ curves is related to the strengthening of pinning potential with decreasing temperature, which yields higher stopping forces on the vortex lattice. The value of the critical current increases as temperature decreases: values for H=32 Oe are around 25 kA.cm$^{-2}$ (1 mA in our bridge) at T=0.99$T_c$ and 100 kA.cm$^{-2}$ (4 mA) at T=0.98$T_c$,. These are typical values of critical current in this kind of patterned Nb films.[12,18,19]. It is remarkable that the effect of temperature on our experimental ratchet is very similar to the one produced by disorder in the array of asymmetric pinning centers on the simulated ratchet effect calculated by Zhu et al..[16] The shape of the simulated curves for higher disorder (Figure 11 of Ref. 16) mimics our experimental curves for higher temperatures, while the steps and sharp asymmetric shapes experimentally observed at low temperatures are similar to simulated curves for low disorder in the array. According to this fact, the enhancement of the ratchet effect due to the decrease of temperature is similar to the one observed in simulations for higher array order. On the other hand, the ratchet effect vanishes when the temperature, or the array disorder, increases. These features in common should be explored deeper.

Finally, the effects of increasing the applied magnetic field and its temperature dependence have been studied. Increasing the applied magnetic, which yields interstitial vortices (vortices not pinned in Ni triangles), produces a striking effect.[2] The polarity of the ratchet signal changes when going from three vortices per unit cell (everyone of the vortices in the triangles) to four vortices per unit cell (the first interstitial vortex appears). The reversed polarity of the ratchet is explained because the interstitial vortices feel a weaker and reverse ratchet potential than the pinned vortices, since the asymmetry is induced for those by vortex-lattice interactions. The negative signal appears at low drives. The interstitials move first, they need smaller applied force than the pinned vortices to yield a dc voltage.

Fig. 4 shows this behavior for two different temperatures. Again we can notice that decreasing the temperature the ratchet effect is enhanced for both types of vortices, although the distortion on the signal shape only appears in the case of pinned vortices, since they are the vortices in the ratchet traps, and the interstitial vortices feel the weaker ratchet effect through the vortex lattice interaction.

In closing, applying an alternating current to a Nb film growth on Ni arrays with asymmetric pinning potentials yields net vortex motion, whose sense is governed by the asymmetry of the pattern. The most outstanding experimental facts of this superconducting experimental ratchet are:

i) The origin of the ratchet effect is the asymmetry of the potentials that assemble the array.

ii) The ratchet effect is monotonically smoothed increasing the temperature.

iii) At constant temperature and driving force, the ratchet effect reverses and decreases amplitude increasing the applied magnetic field. That is, the number of vortices per unit cell of the array and the amplitude of the *ac* driving current govern the polarity and amplitude of the *dc* ratchet voltage.

iv) At constant temperature and applied magnetic field, the ratchet effect vanishes increasing the driving force on the vortex-lattice.

We want to thank Spanish CICYT MAT2002-04543 and R. Areces Foundation. EMG wants to thank Ministerio de Educación y Ciencia for a Ramon y Cajal contract.

Figure captions:

**Figure 1: (a)** $V_{dc}$ output versus *ac* input amplitude $I_{ac}$ for sample A, with current injected along *x*-axis, at T=0.989$T_c$ and applied magnetic field H=32 Oe (*n*=1). **Inset:** Sketch of the array of triangles and their sizes for sample A. **(b)** $V_{dc}$ output versus *ac* input amplitude $I_{ac}$ for sample B, with current injected along *x*-axis at T=0.984$T_c$ and applied magnetic field H=54 Oe (*n*=1). **Upper inset:** $V_{dc}$ output versus *ac* input amplitude $I_{ac}$ for sample B, with current injected along the *y*-axis, at T=0.984$T_c$ and applied field H=54 Oe (*n*=1). **Lower inset:** Sketch of the array of triangles and their sizes for sample B.

**Figure 2: (Upper panel)** *ac* ratchet effect $V_{dc}(I_{ac})$ for sample A (black dots), and *dc* ratchet effect obtained by subtracting $V(I_+)-|V(I_-)|$ (white dots), at applied magnetic field H=64 Oe, and T=0.995$T_c$ **(Lower panel)** $V(I_+)$ and $|V(I_{dc-})|$ at H=64 Oe and T=0.995$T_c$.

**Figure 3:** $V_{dc}$ output versus $J_{ac}$ amplitude input (or equivalent net velocity *v* versus *ac* force amplitude $F_L$) for sample A, with *J* injected along the *x*-axis, at applied magnetic field H=32 Oe (*n*=1) and temperatures 1: T=8.300 K, 2: T=8.265 K, 3: T=8.250 K, 4: T=8.214 and 5: T=8.170 K. Sample $T_c$=8.340 K. **Inset:** Maximum *dc* rectification $V_{dc}^{max}$ and *ac* current input at which this is achieved $J_{as}^{max}$ as a function of the reduced temperature T/$T_c$, for sample A, with *J* injected along *x*-axis, at applied magnetic field H=32 Oe (*n*=1).

**Figure 4: (Upper panel)** For sample A, net velocity of the vortex lattice *v* as a function of the *ac* Lorentz force amplitude $F_L$, for a frequency ω=10kHz , T=0.979$T_c$ and different numbers of vortices per unit cell *n*. **(Lower panel)** For sample A, net velocity

of the vortex lattice $v$ as a function of the ac Lorentz force amplitude $F_L$, for a frequency $\omega$=10kHz , T=0.989$T_c$ and different numbers of vortices per unit cell $n$.

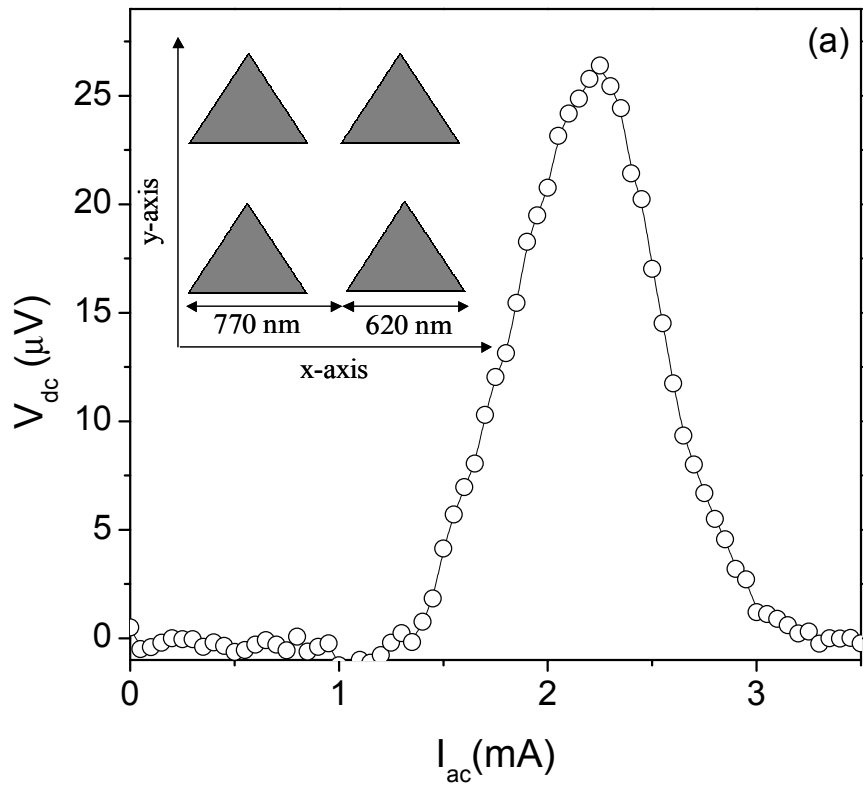

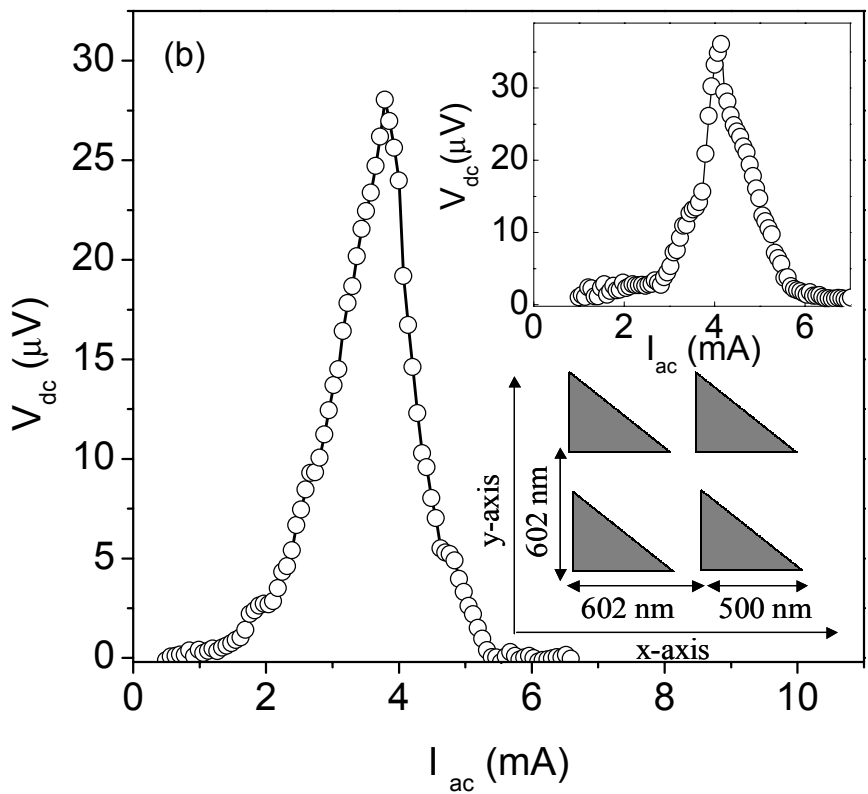

Figure 1
J.E. Villegas *et al.*

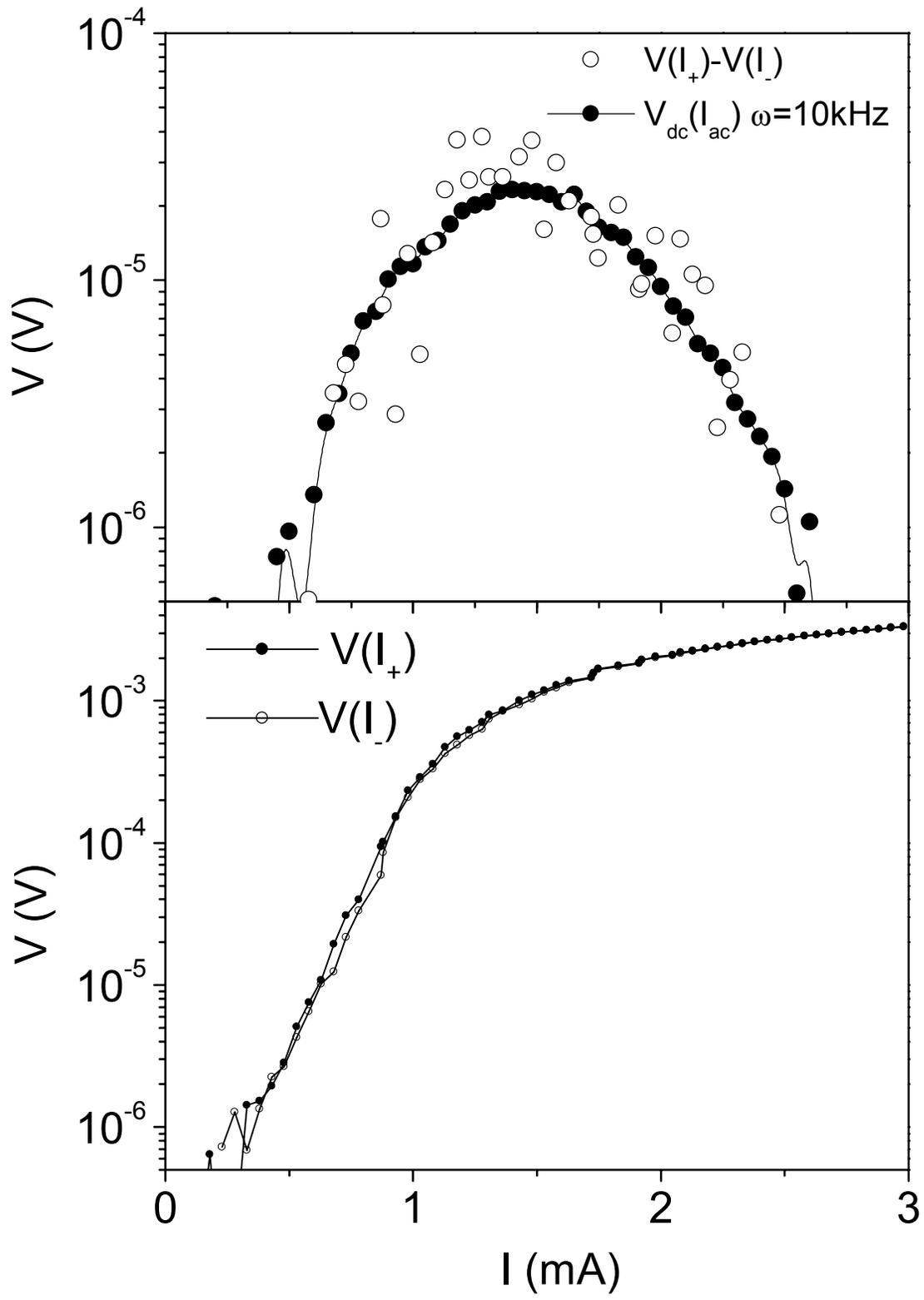

Figure 2
J.E. Villegas *et al.*

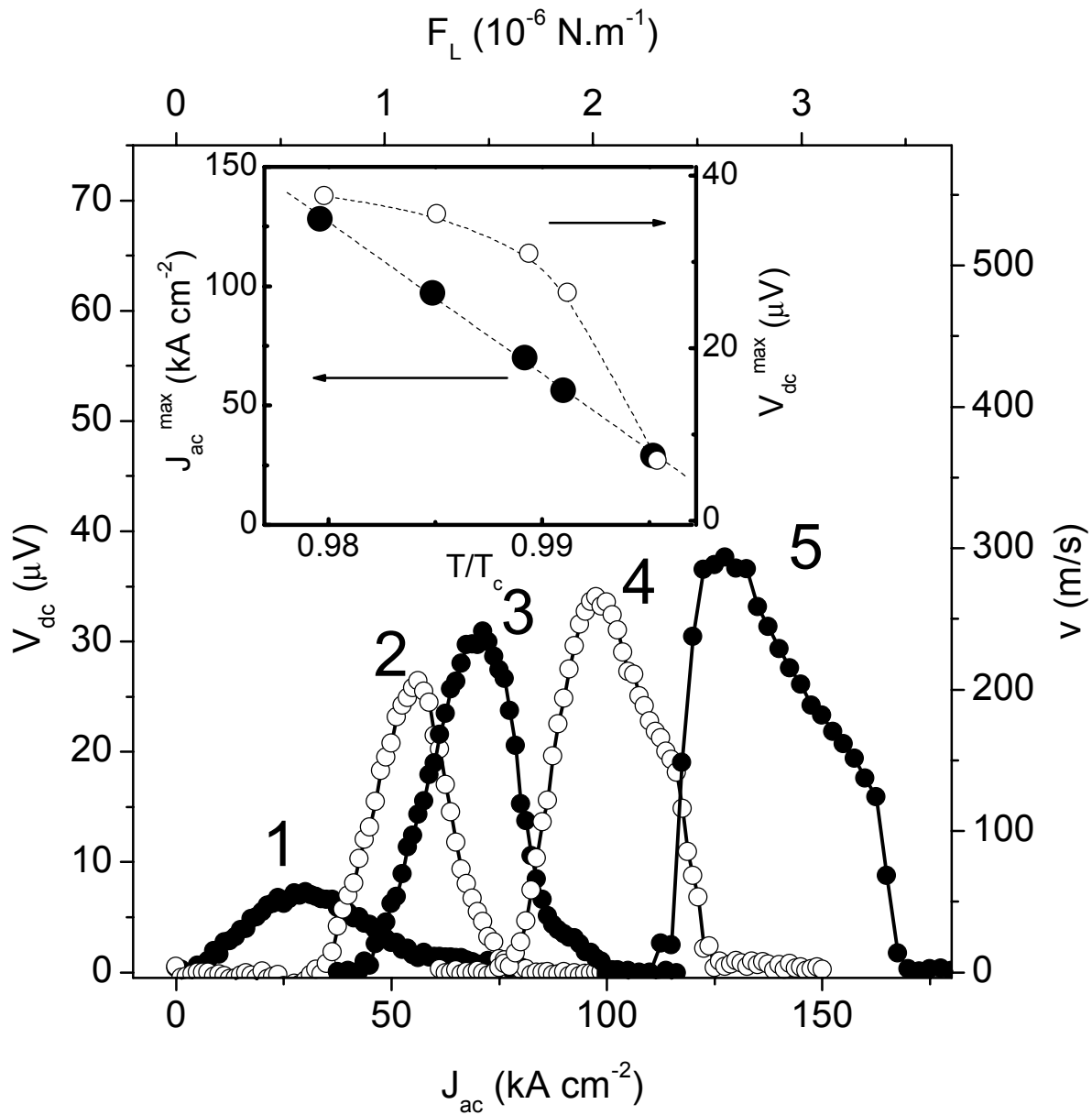

Figure 3
J.E. Villegas *et al.*

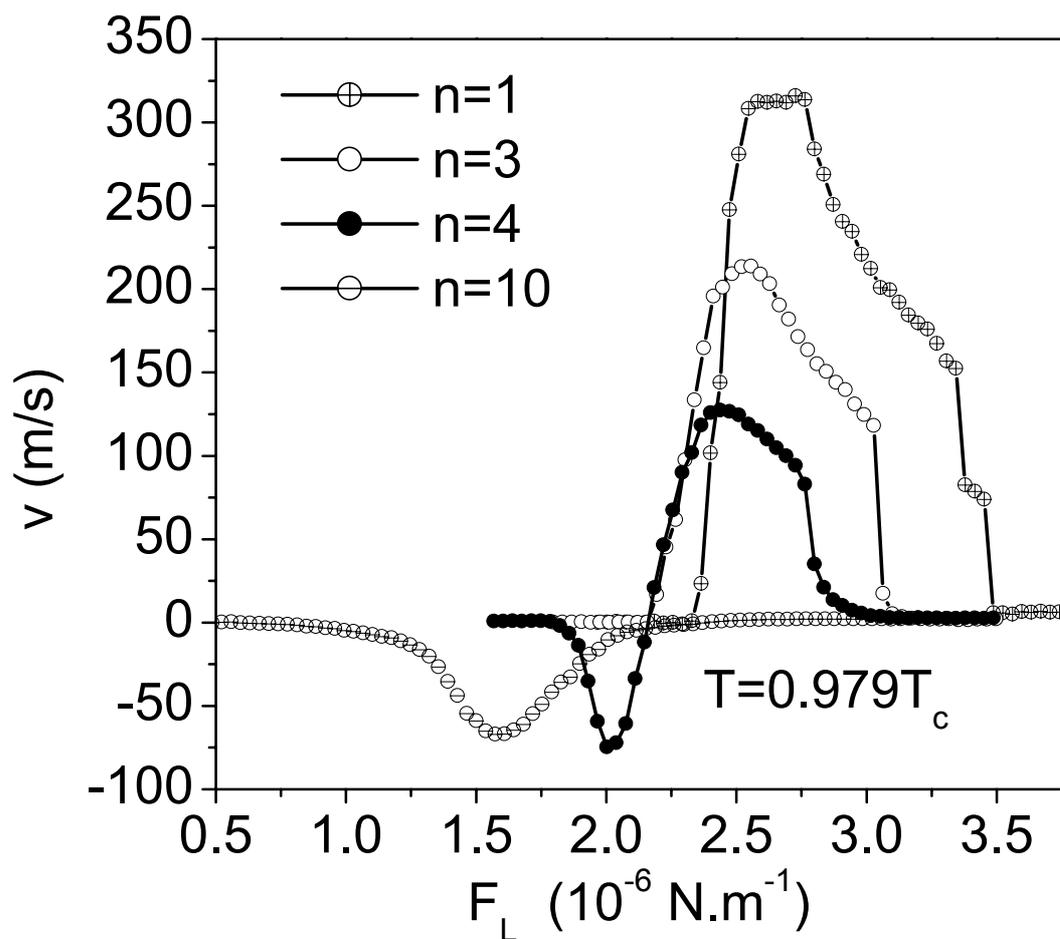
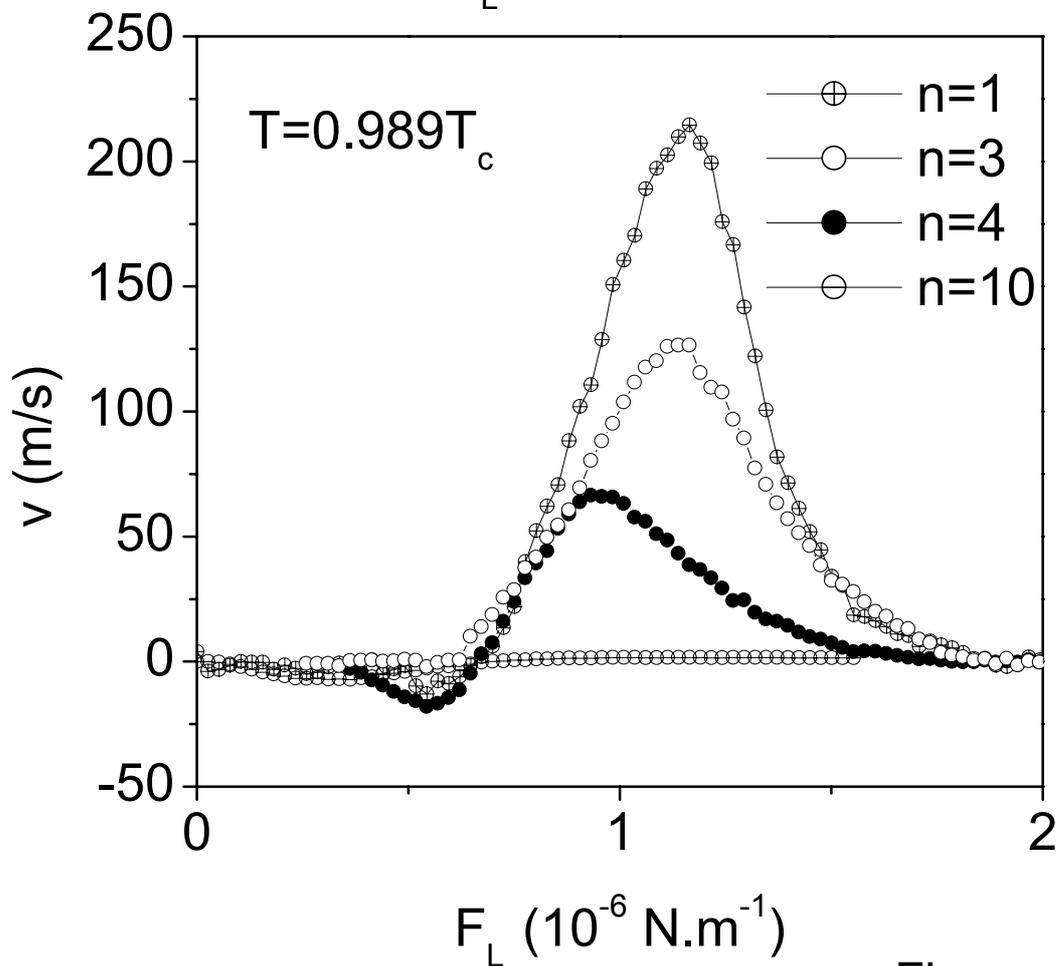

Figure 4
J.E. Villegas *et al.*